\documentclass[pra,showpacs,floatfix]{revtex4}
\usepackage{amssymb}
\usepackage{graphicx}

\newcommand{\beq}{\begin{equation}}
\newcommand{\eeq}{\end{equation}}
\newcommand{\beqa}{\begin{eqnarray}}
\newcommand{\eeqa}{\end{eqnarray}}
\newcommand{\ba}{\begin{array}}
\newcommand{\ea}{\end{array}}

\begin{document}

\title{Fundamental solitons in discrete lattices with a delayed nonlinear
response}
\author{A. Maluckov$^{1}$, Lj. Had\v{z}ievski$^{2}$, and B. A. Malomed$^{3}$}
\affiliation{$^1$ Faculty of Sciences and Mathematics, University of Ni\v s, P.O. Box
224, 18001 Ni\v s, Serbia \\
$^2$ Vin\v ca Institute of Nuclear Sciences, P.O. Box 522,11001 Belgrade,
Serbia \\
$^3$ Department of Physical Electronics, School of Electrical Engineering,
Faculty of Engineering, Tel Aviv University, Tel Aviv 69978, Israel}

\begin{abstract}
The formation of unstaggered localized modes in dynamical lattices can be
supported by the interplay of discreteness and nonlinearity with a finite
relaxation time. In rapidly responding nonlinear media, on-site discrete
solitons are stable, and their broad inter-site counterparts are marginally
stable, featuring a virtually vanishing real instability eigenvalue. The
solitons become unstable in the case of the slowly relaxing nonlinearity.
The character of the instability alters with the increase of the delay time,
which leads to a change in the dynamics of unstable discrete solitons. They
form robust localized breathers in rapidly relaxing media, and decay into
oscillatory diffractive pattern in the lattices with a slow nonlinear
response. Marginally stable solitons can freely move across the lattice.
\end{abstract}

\maketitle

\textbf{The delay of the medium's response is unavoidable in natural
phenomena and technology. In this respect, it is relevant to mention the
cellular physiology, genetics, signal processing, transport of impulses
through neural networks, the propagation of light signals in optical
resonator systems and cavities, and so on. The delayed response may
significantly affect many phenomena in these fields. In particular,
essential issues, which are addressed in this paper, are the existence,
stability and dynamics of localized modes in dynamical lattices with the
non-instantaneous nonlinear response. The lattice, which may describe an
array of nonlinear optical waveguides, is modeled by the time-delayed
discrete nonlinear Schr\"{o}dinger equation. The localized modes in this
model are explored using a properly modified delay-differential equation
solver package DDE-BIFTOOL v. 2.00. We show that the discrete bright on-site
and inter-site solitons (so classified according to the location of their
center) can be created by the modulational instability of continuous waves,
which occurs in the same parameter region as in the lattice with the
instantaneous nonlinearity. However, the stability of the solitons is
affected by the delayed response in three different ways: (i) the solitons
are destroyed when the delay time exceeds a certain critical value; (ii) in
the case of the fast nonlinear response (short temporal delay), the
instability growth rate of inter-site solitons decreases in comparison with
the instantaneous model; (iii) the instability type changes for intermediate
values of the delay time. The inter-site and on-site solitons with very
close values of the power in the fast-responding media evolve into localized
breathing modes when their amplitude is slightly perturbed. They can also be
transformed into moving localized modes by the application of a kick.
However, larger values of the delay time enhance the temporal correlation
between time-distant events, which results in bringing moving solitons to a
halt.}

\section{Introduction}

Usually, whenever an irregular, unpredictable change occurs in an
engineering device, the need arises to take action in order to make the
perturbation predictable or suppress it altogether. An solution of this
problem is based on a delayed feedback, in the form of a signal proportional
to the difference between the current value of some parameter of the
system's state and its value in the past. Taking the time delay equal to the
period of the unwanted random oscillations, the evolution may be made
stable, with the appropriate choice of the feedback strength \cite%
{balanov,ahlborn,netw}.

In particular, systems built as recurrent neural networks have found many
applications to signal and image processing, pattern recognition,
realization of associative memories, solving certain optimization problems,
and in other fields. Because delays of integration and communication are
unavoidable, often causing an instability of the systems, attention has been
focused on the stability of neural networks with a temporal delay \cite%
{netw1,netw2,netw3}. Time-delayed systems also play a significant role in
modeling living cells, genetics, etc. \cite{cellular,genetics,oscill}.

In the field of optics, a strong delay is demonstrated by coupled-resonator
optical waveguides (CROWs), which are built as side-coupled densely packed
chains of spherical cavities \cite{CROW}. It may happen that a light wave
spends much more time circulating in each cavity than hopping between them,
which makes the group velocity of the light transmission very low. Because
the time delay induced by the sojourn of the light signal in the resonator
strongly depends on the detuning of the signal from the resonator's
eigenfrequency, CROWs feature the group-velocity dispersion which is much
stronger than in conventional materials \cite{Kobi1,slow}. Applications
suggested by such settings include various slow-light phenomena,
transmission lines with a variable optical delay, clean pulse compression on
a chip without formation of a pedestal, etc. \cite%
{Kobi2,opt,opt1,lasers,breathers}.

The above-mentioned examples demonstrate that the delay of the medium's
response may significantly affect various fundamental phenomena and
applications. In this regard, an essential issue is the dynamics of \emph{%
localized modes} in media with the non-instantaneous nonlinear response. In
particular, it was demonstrated that the relaxation of the nonlinearity has
a strong influence on the wave-packet dynamics \cite{breathers,moura}.

The objective of the present work is to study the existence,
stability and dynamics of localized modes in one-dimensional
discrete lattices with the on-site non-instantaneous nonlinear
response. We find that the solitons in this case can be formed by
the modulation instability (MI), similarly to the lattice with the
instantaneous response. We explore the dependence of the stability
and dynamical properties of fundamental unstaggered localized
modes in the lattice on the response time ("unstaggered" means
that the sign of the lattice field does not jump between adjacent
sites \cite{Panos}).

It is relevant to mention that the stability and related dynamical
properties of the time-delayed systems have been studied in some detail
since the pioneering work \cite{old}, where several stability criteria for
discrete delayed systems were derived, in a form independent of the value of
the delay time. A rigorous stability analysis has been performed for simple
linear and nonlinear systems with the temporal delay, that were modeled by
differential-difference equations \cite{amann,numst}, primarily with the
objective to predict a possibility of controlling the system through the
time-delayed feedback. In addition, the stability and bifurcations of fixed
points have been numerically investigated for a few simple nonlinear models
of the neural activity \cite{num}. In this work, we looked for numerical
solutions for the stationary solitons in the TD-DNLSE by means of modified
dde-solver tools \cite{dde}, while the stability was analyzed using a
properly modified MATLAB program package DDE-BIFTOOL v. 2.00 \cite%
{ddebiftool,num}.

The rest of the paper is structured as follows. The model is formulated in
Section II, following Ref. \cite{moura}, where it was introduced for
electron-photon systems featuring a nonadiabatic interaction. The MI is also
considered in Section II. The main results concerning the stability and
dynamical behavior of the discrete solitons are presented in Section III.
The paper is concluded by Section IV.

\section{The modulation instability in lattices with the non-instantaneous
nonlinearity}

In solid-state media, the transmission of electron wave packets through
chains with a nonadiabatic on-site electron-phonon interaction may be
modeled by the \textit{time-delayed discrete nonlinear Schr\"{o}dinger
equation} (TD-DNLSE), which was proposed, in this context, in Ref. \cite%
{moura}:
\begin{equation}
i\frac{d\psi _{n}(t)}{dt}+C\left[ \psi _{n+1}(t)+\psi _{n-1}(t)-2\psi _{n}(t)%
\right] -\kappa |\psi _{n}(t-\tau )|^{2}\psi _{n}(t)=0,  \label{ddf}
\end{equation}%
where $n$ is the discrete coordinate, $\tau $ is the delay time, $C>0$ is
the coupling constant, and $\kappa $ is the nonlinearity parameter which
determines the sign of the nonlinearity, $\kappa =-1$ and $+1$ corresponding
to the self-attraction and self-repulsion, respectively. In the simplest
approximation, this model may be applied to the description, in the temporal
domain, of optical systems in the form of the above-mentioned CROWs. In the
next section, we focus on the case of $\kappa =-1$, when the unstaggered
solitons exist in the model with the instantaneous nonlinearity.

The only dynamical invariant which remains conserved in the presence of the
non-instantaneous nonlinearity is the total norm (alias energy, in terms of
the CROW-like optical models):
\begin{equation}
P(t)=\sum_{n=1}^{N}|\psi _{n}(t)|^{2}.  \label{P}
\end{equation}%
Below, it will be used as the main characteristic of localized modes.

As is well known \cite{nenash,Panos,nash1}, the instantaneous DNLSE exhibits
the MI of continuous-wave (CW) states in the lattice, which is a mechanism
for the creation of discrete solitons. The MI is a result of the competition
between the discreteness and the nonlinearity. Here we analyze how the
delayed response affects the MI.

\begin{figure}[tbp]
\center\includegraphics [width=7cm]{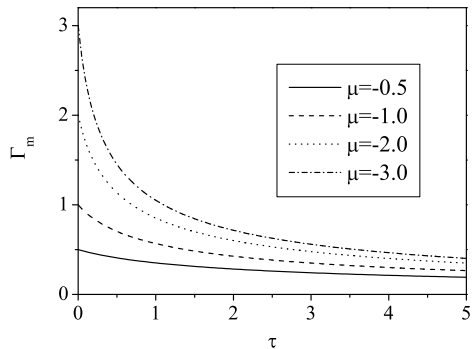} \caption{The MI
growth rate, $\Gamma _{m}$, corresponding to the fastest growing
modulation mode, versus the delay time, $\protect\tau $, for a few
fixed values of frequency $\protect\mu $ which determines the
amplitude of the unstable CW state (\protect\ref{CW}).}
\label{fig1}
\end{figure}

Equation (\ref{ddf}) gives rise to the CW solution, $\psi _{n}(t)=A\exp {%
(-i\mu t)}$, with frequency $\mu $ and constant amplitude $A=\sqrt{\mu
/\kappa }$, in the regions of $\mu <0$ for $\kappa =-1$, and $\mu >0$ for $%
\kappa =+1$. The MI can be investigated, in the framework of the
linear-stability analysis, by adding a small complex perturbation to the CW
solution,
\begin{equation}
\psi _{n}(t)=\left[ A+\delta u_{n}(t)\right] \exp {(-i\mu t)}  \label{CW}
\end{equation}%
\cite{nenash,Panos,nash1}. Looking for solutions for the perturbations in
the form of $\delta u_{n}(t)\sim \exp {(iqn)}\exp {(\Gamma t)}$, where $q$
is the wavenumber and $\mathrm{Re}\{\Gamma \}$ is the growth rate of the
perturbation, we obtain the dispersion relation for the lattice with the
delayed response, in the form of the following transcendental equation:
\begin{equation}
\Gamma ^{2}=-8C\sin ^{2}\left( {q/2}\right) \left[ 2C\sin ^{2}\left( {q/2}%
\right) +\mu \exp {(-\Gamma \tau )}\right] .  \label{disp}
\end{equation}%
The CW solution is unstable if Eq. (\ref{disp}) gives rise, for some
wavenumbers $q$, to real positive $\Gamma $. Obviously, this is possible for
$\mu <0$, which coincides with the MI condition at $\tau =0$ (recall the CW
with $\mu <0$ exists for $\kappa =-1$, i.e., in the case of the
self-attractive nonlinearity). However, the time delay affects the magnitude
of $\Gamma $. The expression for the growth rate corresponding to the
fastest-growing instability mode can be obtained from Eq. (\ref{disp}):
\begin{equation}
\Gamma _{m}=W(-\mu \tau )/\tau ,  \label{gr}
\end{equation}%
where $W(x)$ is the Lambert's $W$ function \cite{lambert}, defined as a
solution to equation $x=We^{W}$. The dependence of $\Gamma _{m}$ on $\tau $,
which was numerically calculated for different values of $\mu $, is
presented in Fig. \ref{fig1}. It clearly shows that the increase of the
delay time leads to a \emph{decrease} of $\Gamma _{m}$.

\section{The stability and dynamics of fundamental solitons}

\subsection{The general approach}

The development of the MI may split the CW into an array of solitons. The
stationary solution for a discrete soliton with frequency $\mu $ can be
looked for as $\psi _{n}(t)=\phi _{n}\exp {(-i\mu t)}$, where $\phi _{n}$ is
a real lattice field, subject to conditions $\phi _{n\rightarrow \pm \infty
}\rightarrow 0$, which satisfies the following equation:
\begin{equation}
\mu \phi _{n}+C(\phi _{n+1}+\phi _{n-1}-2\phi _{n})+\phi _{n}{}^{3}=0,
\label{stat}
\end{equation}%
where we set $\kappa =-1$, as said above, because unstaggered solitons do
not exist in the opposite case. Equation (\ref{stat}) has the same form as
in the case of $\tau =0$, hence the shape of the stationary solitons (unlike
their stability) is independent on the non-instantaneous character of the
nonlinearity. For this reason, examples of stationary discrete solitons are
not displayed here (they are identical to those in the standard DNLSE with
the instantaneous self-attractive nonlinearity \cite{Panos}).

The stability of the solitons was analyzed following the standard
linearization procedure. To this end, a small complex perturbation is added
to the soliton,
\begin{equation}
\psi _{n}(t)=\left[ \phi _{n}+\delta a_{n}(t)+i\delta b_{n}(t)\right] \exp {%
(-i\mu t)},  \label{pert}
\end{equation}%
where $\delta a_{n}$ and $\delta b_{n}$ are real functions, and $\mu $ is
the same frequency of the unperturbed soliton which was defined above.
Substituting expression (\ref{pert}) into Eq. (\ref{ddf}) and performing the
linearization, a system of equations for the small perturbations is
obtained:
\begin{eqnarray}
\frac{d}{dt}\delta a_{n}(t) &=&\left[ (2C-\mu )-\phi _{n}^{2}\right] \delta
b_{n}(t)-C\left[ \delta b_{n+1}(t)+\delta b_{n-1}(t)\right] , \\
\frac{d}{dt}\delta b_{n}(t) &=&-\left[ (2C-\mu )-\phi _{n}^{2}\right] \delta
a_{n}(t)+C(\delta a_{n+1}(t)+\delta a_{n-1}(t))+2\phi _{n}^{2}\delta
a_{n}(t-\tau ),
\end{eqnarray}%
which can be rewritten in the matrix form:
\begin{equation}
\frac{d}{dt}U(t)=M_{1}U(t)+M_{2}U(t-\tau ),  \label{matrix}
\end{equation}%
$U(t)$ being the column transpose to $\left\{ \delta a_{1}(t)...\delta
a_{N}(t)\delta b_{1}(t)...\delta b_{N}(t)\right\} $, and $M_{1,2}$ are the
corresponding $2N\times 2N$ matrices.

A solution of Eq. (\ref{matrix}) gives rise to $2N$ stability eigenvalues
(EVs), the soliton being unstable if there are EVs with nonzero real parts.
The instability is exponential or oscillatory if, respectively, the
corresponding EVs are real or complex. In the area-preserving models
(Hamiltonian lattices, which is the case only for $\tau =0$), real EVs
always appear in pairs of positive and negative ones with equal absolute
values, while complex EVs appear in quartets featuring different
combinations of the signs in the front of the real and imaginary parts \cite%
{Panos}. Obviously, the presence of the time delay in Eq. (\ref{matrix}) may
alter the EVs and affect the stability of the discrete solitons.

Our main aim is to study the stability of stationary unstaggered fundamental
solitons in the lattice system with the finite response time. We consider
the lattice subject to periodic boundary conditions, with the soliton's
center placed either at the site of the lattice (\textit{on-site} solitons),
or between two sites (\textit{inter-site} solitons). Results presented below
were obtained for the lattices with $N=100$ or $101$ sites, for the
inter-site and on-site modes, respectively. Values of the delay time were
taken in the interval of $0<\tau <5$.

\begin{figure}[tbp]
\center\includegraphics [width=6cm]{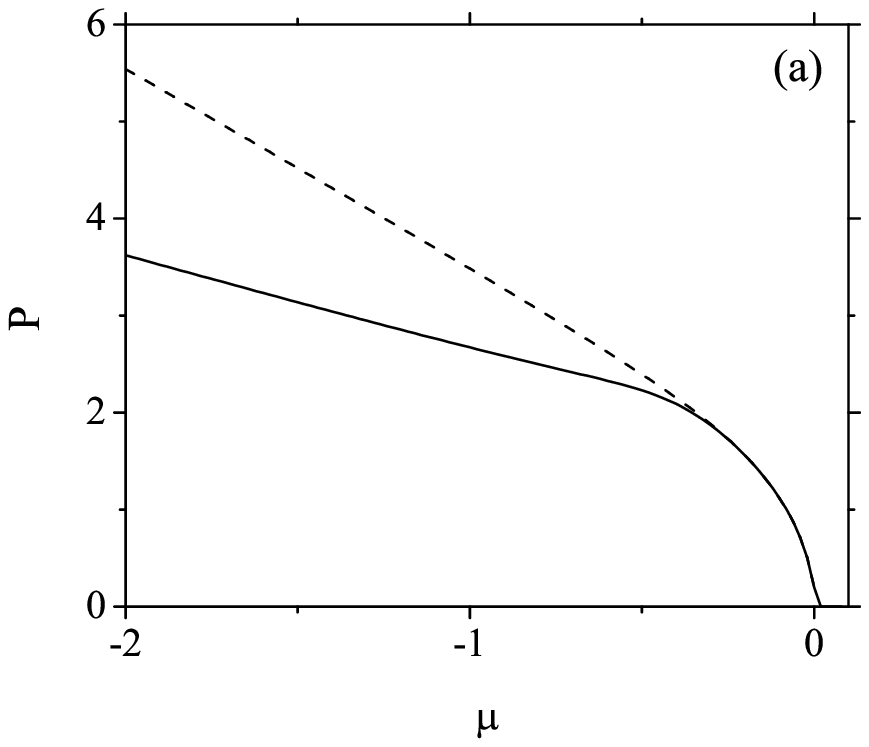}
\includegraphics
[width=6cm]{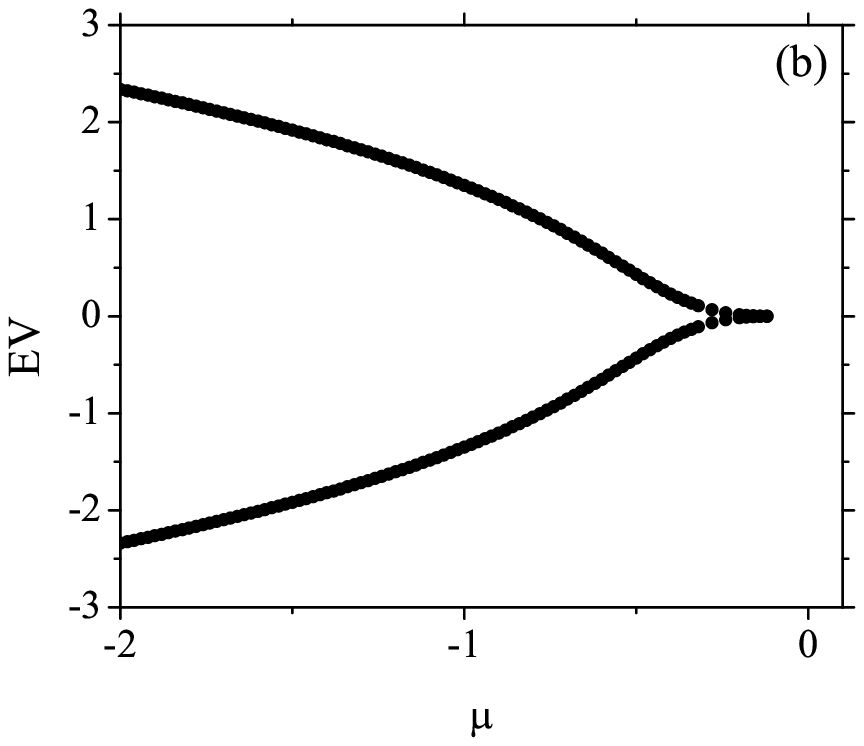} \caption{(a) Norm versus $\protect\mu $ for
on-site (solid line) and inter-site (dashed line) stationary
fundamental solitons in the lattice with $C=0.8$. The EV
(eigenvalue) spectrum for the inter-site soliton in the absence of
the temporal delay, $\protect\tau =0$, includes a pair of pure
real EVs, as shown in (b), for all $\protect\mu $, except in an
area close to $\protect\mu =0$, where the unstable EV practically
vanishes, and the on-site counterpart of the inter-site soliton,
with the same norm, can be found, cf. panel (a). } \label{fig2}
\end{figure}

The lattices with the instantaneous nonlinear response support unstaggered
fundamental solitons of the two above-mentioned types, on-site and
inter-site ones \cite{nenash,Panos,nash1}. In terms of the corresponding EVs
of small perturbations, the on-site solitons are stable in the Kerr media
with the attractive cubic nonlinearity, while the inter-site solitons are
unstable. A specific situation occurs at $\mu \rightarrow -0$, where the
unstable inter-site solitons can find on-site counterparts with very close
values of the norm (\ref{P}), see Fig. \ref{fig2}(a). In that case, the
solitons of both types may be called \textit{marginally stable }\cite{nash1}%
, as the instability growth rate of the inter-site ones virtually vanishes,
see Fig. \ref{fig2}(b), and the dynamical behavior of the on-site and
inter-site solitons with equal norms is nearly identical (under the action
of perturbations, they become breathers, see below). These solitons feature
large widths and small amplitudes.

Being slightly perturbed, the unstable inter-site soliton spontaneously
shifts by half a lattice spacing, to transform itself into a stable on-site
counterpart with the same norm. In the region of the coexistence of
inter-site and on-site solitons with nearly equal norms, a kicked soliton
may move along the lattice, by hopping between the on-site and inter-site
configurations, overcoming the Peierls-Nabarro potential barrier. Through
this mechanism, the barrier effectively vanishes for the marginally stable
solitons, making them fully mobile objects.

\subsection{Inter-site solitons}

\begin{figure}[hh]
\center\includegraphics [width=6cm]{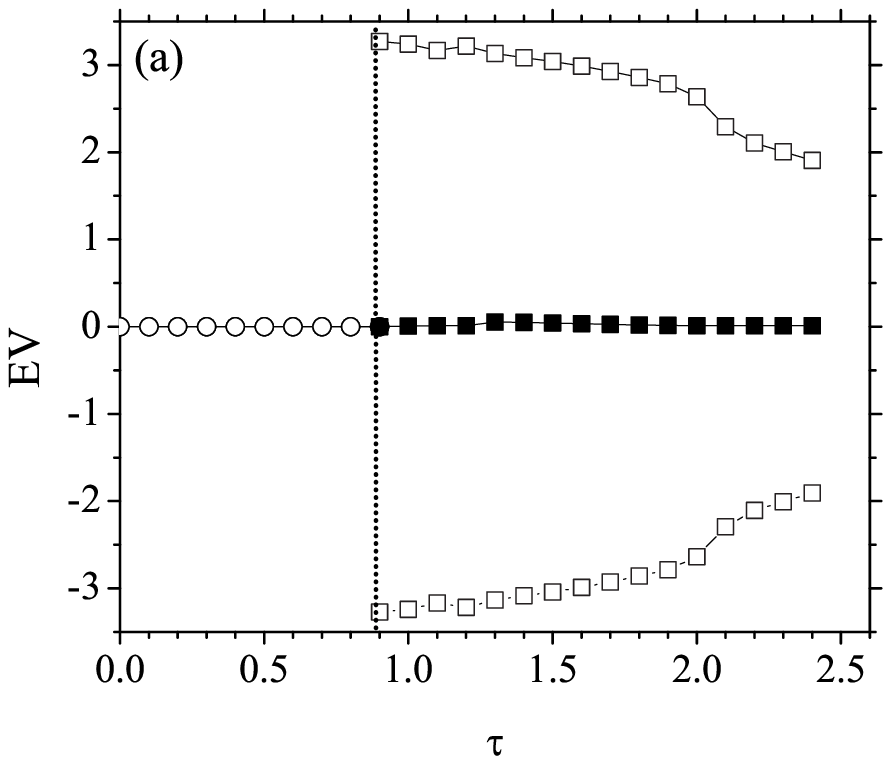}
\includegraphics
[width=5.5cm]{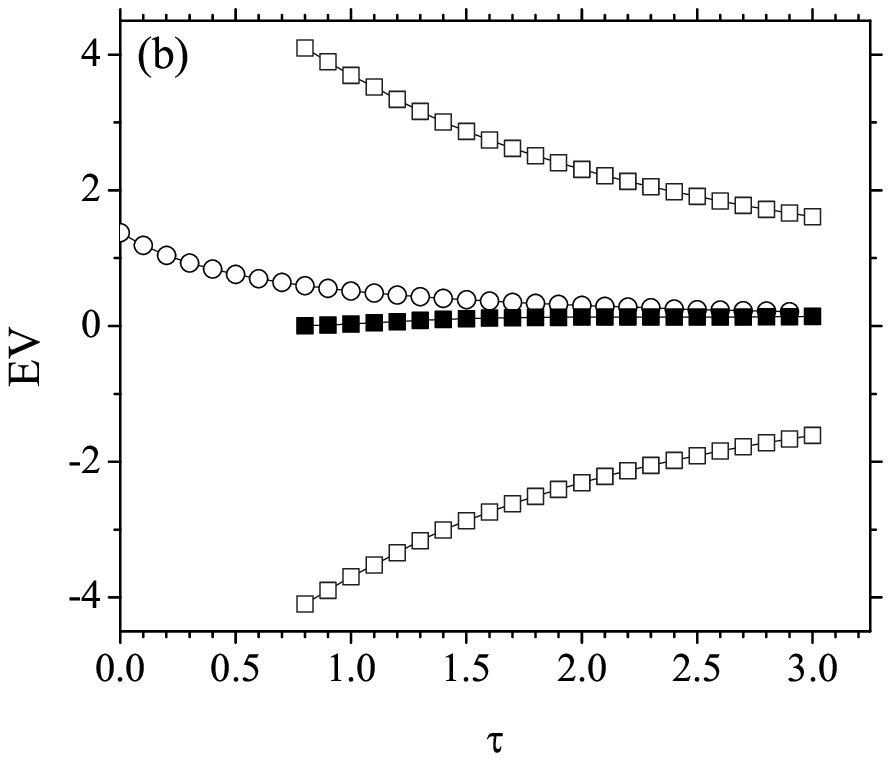} \caption{The EVs (eigenvalues) with the
largest real part are shown, as functions of the delay time,
$\protect\tau $, for inter-site solitons: (a) at $\protect\mu
=-0.22$, and (b) at $\protect\mu =-1.02$. The lattice coupling
constant is $C=0.8$. Dark and white squares denote real and
imaginary parts of the complex EVs. In panel (a), the dotted line
separates the region of the marginally stable inter-site solitons
from that where unstable complex EVs emerge. In panel (b), the
dotted line separates an area of the exponential instability (left
region - circles) from that with many complex EVs. Only complex
EVs with the largest real part are plotted (showing the entire set
of the EVs would make the plots messy). In (b), the
pure real EV and the real part of the complex EV become comparable at $%
\protect\tau >1.5$.} \label{fig3}
\end{figure}

Marginally stable inter-site solitons, which we again define as
those with virtually vanishing real eigenvalues, and with on-site
counterparts possessing the same norm, can be found, at relatively
small values of $|\mu | $, in the lattice with a sufficiently
short delay of the nonlinear response, as shown in Fig.
\ref{fig3}(a) for $\mu =-0.22$ and $C=0.8$. All the values of the
instability growth rate in Fig. \ref{fig3}(a) are bounded to
$\mathrm{Re}(\mathrm{EV})\lesssim 0.001$.

\begin{figure}[hh]
\center\includegraphics [width=5cm]{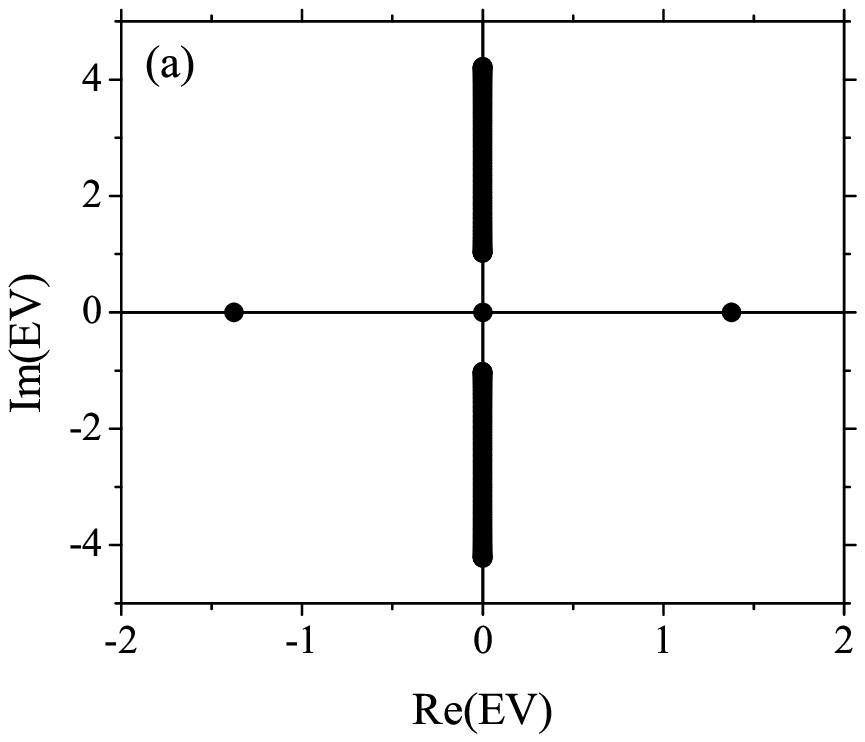}%
\includegraphics
[width=5cm]{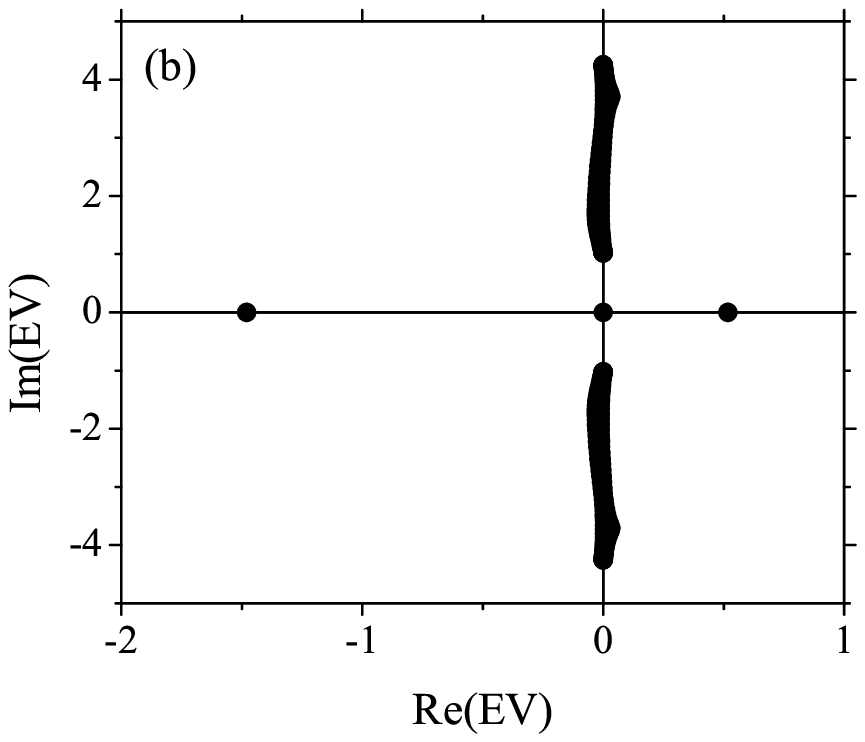}\center
\includegraphics
[width=5cm]{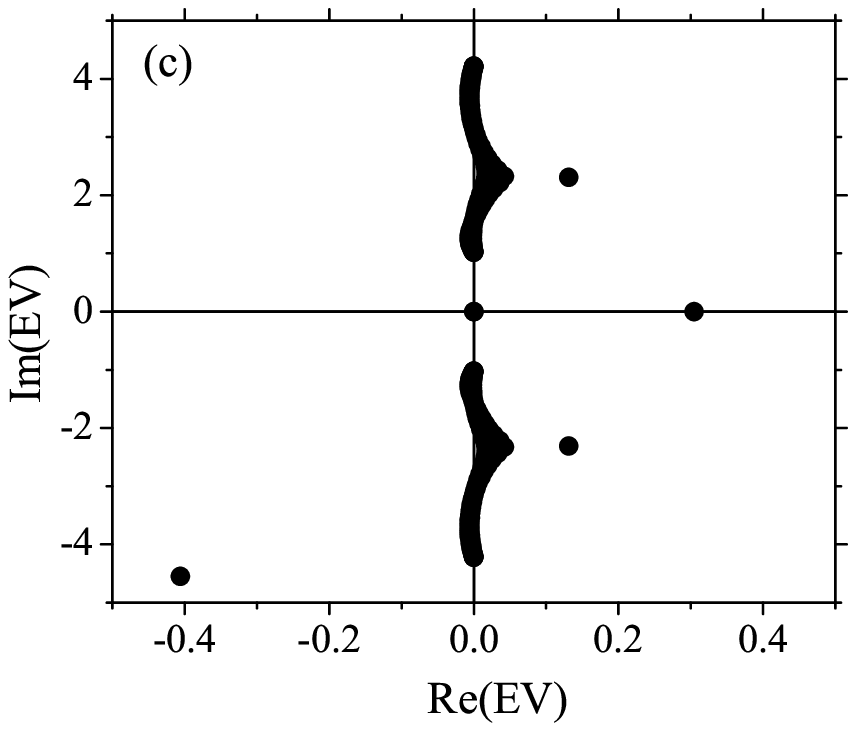}\includegraphics [width=5cm]{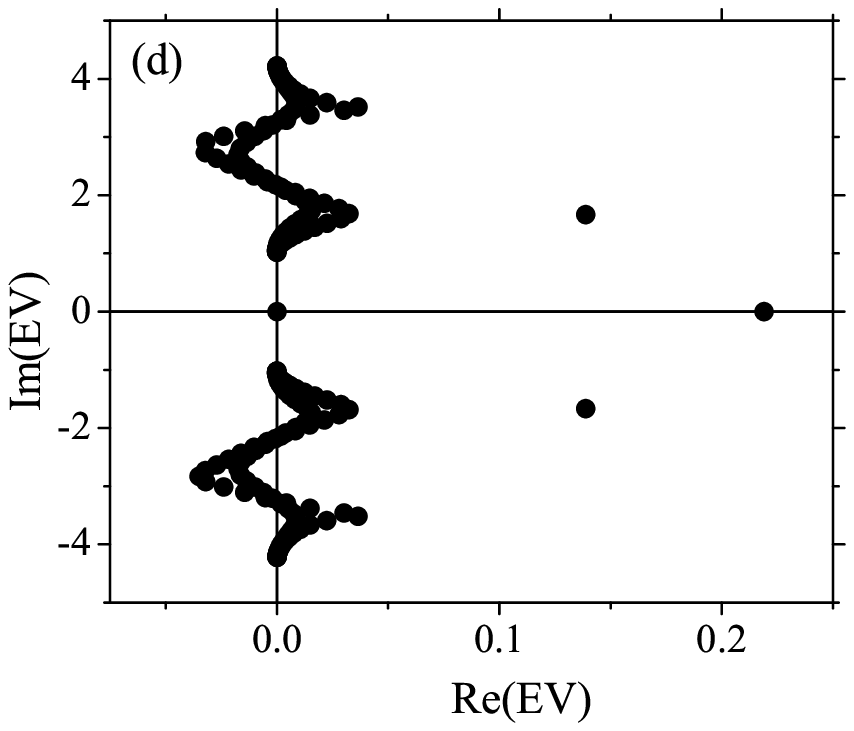}
\caption{The EV spectrum, shown in the form of the EV's imaginary
part
versus the real part, for inter-site solitons. Parameters are $C=0.8$, $%
\protect\mu =-1.02$, and $\protect\tau =0$ (a), $\protect\tau =1$ (b), $%
\protect\tau =2$ (c), and $\protect\tau =3$ (d). All these values
correspond
to particular points in Fig. \protect\ref{fig3}(b). In the case of $\protect%
\tau =0$, the corresponding inter-site soliton is exponentially
unstable [see Fig. \protect\ref{fig3}(b)]. With the increase of
$\protect\tau $, complex EVs with a significant real part start to
emerge, although the real EV remains the dominating one.}
\label{fig4}
\end{figure}

\begin{figure}[hh]
\center\includegraphics [width=12cm]{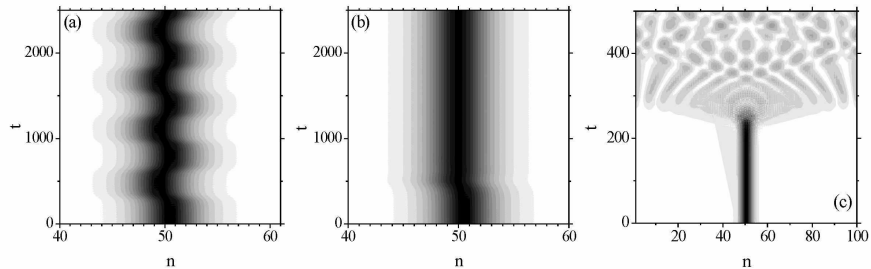}%
\caption{The evolution of a perturbed inter-site soliton with $C=0.8$, $%
\protect\mu =-0.22$. For small $\protect\tau $, it is marginally
stable. Being slightly perturbed, it evolves into a breather at
$\protect\tau =0$ (a), or into an on-site-centered soliton at
$\protect\tau =0.2$ (b). At larger values of the delay time
$\protect\tau $, the inter-site soliton is strongly unstable,
decaying into a diffractive pattern at $\protect\tau =1.2$ (c). }
\label{fig5}
\end{figure}

The instability of the inter-site solitons becomes stronger at larger values
of $|\mu |$ (for more narrow solitons). The unstable solitons develop two
types of instability in the lattice with the non-instantaneous nonlinear
response. In the case of small and medium values of the delay time, $\tau $,
the instability growth rate, represented by the dominant pure real
eigenvalue in Fig. \ref{fig3}(b), is \emph{smaller} than in the case of the
instantaneous nonlinearity. However, the inter-site solitons do not become
truly stable in this case (barring the above-mentioned marginally stable one
found at smaller values of $|\mu |$). As seen in Figs. \ref{fig3} (b) and %
\ref{fig4}, the increase of $\tau $ leads to the creation (starting at $\tau
=0.9$) of a growing number of complex eigenvalues with positive real parts
that account for additional oscillatory instabilities, although the real
eigenvalue remains the largest one.

In the TD-DNLSE, as well as in the DNLSE with the instantaneous
nonlinearity, the long-time evolution of unstable solitons depends on the
initial perturbation. In the case of zero or very small $\tau $, adding a
small-amplitude perturbation to the inter-site mode belonging to an
inter-site/on-site soliton pair with equal norms (i.e., it is a marginally
stable soliton, as defined above), gives rise to a breathing mode, see Fig. %
\ref{fig5}(a). On the contrary, at larger values of $\tau $ (when the
inter-site soliton is no longer marginally stable, being explicitly
unstable), an on-site-centered localized mode appears as a result of the
instability development [Fig.\ref{fig5}(b)]. In the model with still larger $%
\tau $ (in the slowly responding medium), the same unstable inter-site mode
suffers an instability-induced diffraction into an oscillating pattern, as
shown in Fig. \ref{fig5}(c). Diffractive patterns are also a final state of
the evolution of slightly perturbed unstable inter-site modes which do not
have on-site (equal-norm) counterparts.

\subsection{On-site solitons}

\begin{figure}[tbp]
\center\center\includegraphics
[width=5cm]{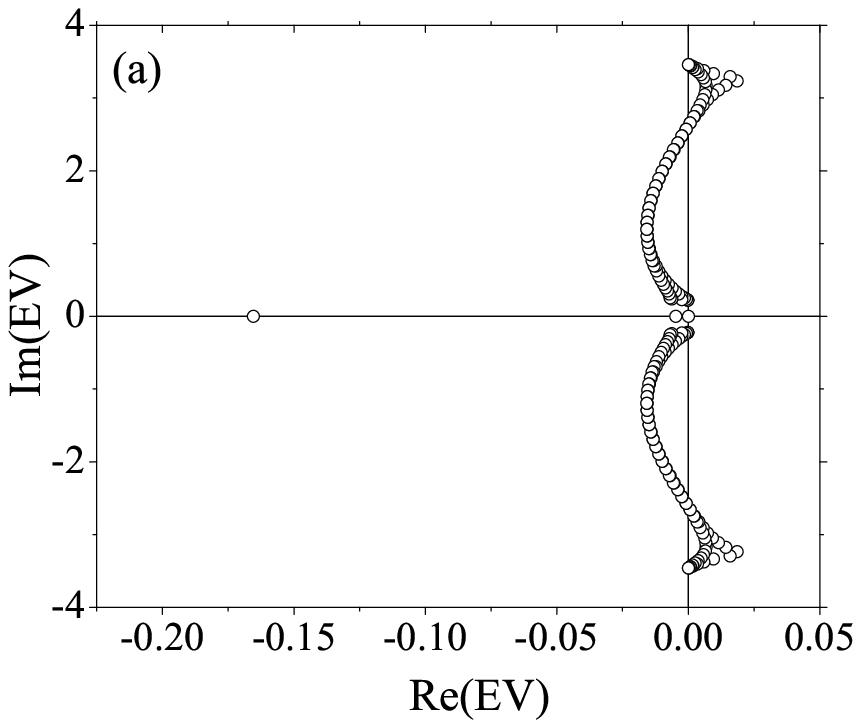}\includegraphics
[width=5cm]{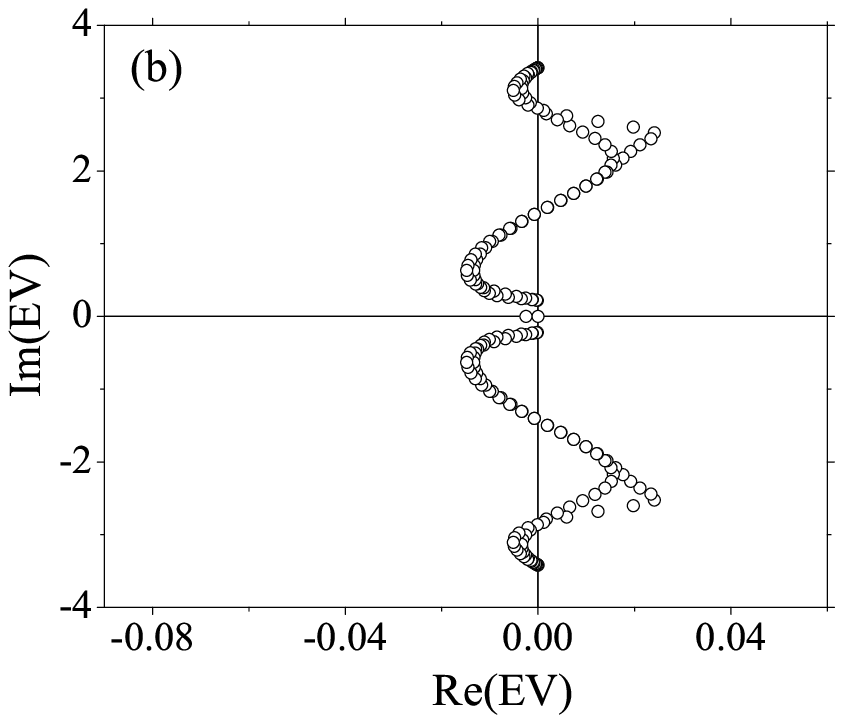}\includegraphics [width=5cm]{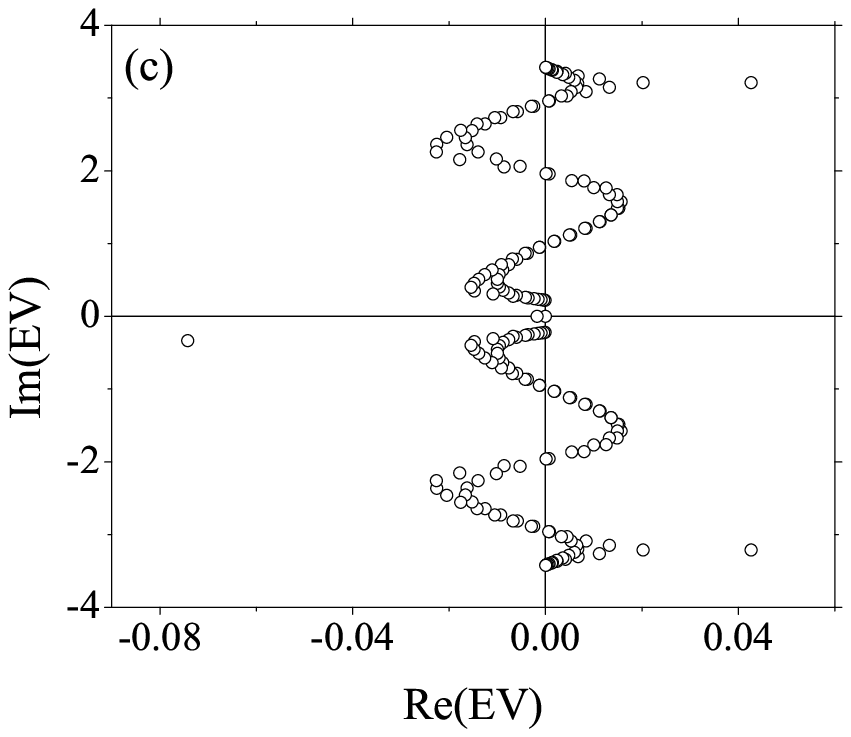}
\caption{The EV spectra for unstable on-site solitons in media
with the non-instantaneous nonlinear response, cf. Fig. 4 for
inter-site solitons. Parameter values are $C=0.8$, $\protect\mu
=-1.02$, and $\protect\tau =1$ (a), $\protect\tau =2$ (b),
$\protect\tau =3$ (c). In the case of a small delay time
$\protect\tau $, the corresponding on-site soliton is stable.}
\label{fig6}
\end{figure}

On-site solitons, which remain stable at sufficiently small values of the
delay time $\tau $, get destabilized with the increase of $\tau $. This is
indicated by a slowly growing pure real EV, corresponding to an eigenmode of
small perturbations which accounts for sharpening of the perturbed on-site
soliton, and by the emergence of a growing number of complex EVs with
positive real parts. For instance, the destabilization of the on-site
solitons is observed at $\tau >0.7$ for $\mu =-1.02$ and $C=0.8$, as shown
in Fig. \ref{fig6}. The perturbed on-site soliton in the slowly responding
media (large $\tau $) decays into a diffractive oscillatory pattern due to
the excitation of unstable oscillatory modes, cf. Fig. 5(c) which shows a
similar outcome of the instability development of the inter-site solitons.

\begin{figure}[tbp]
\center\center\includegraphics [width=7cm]{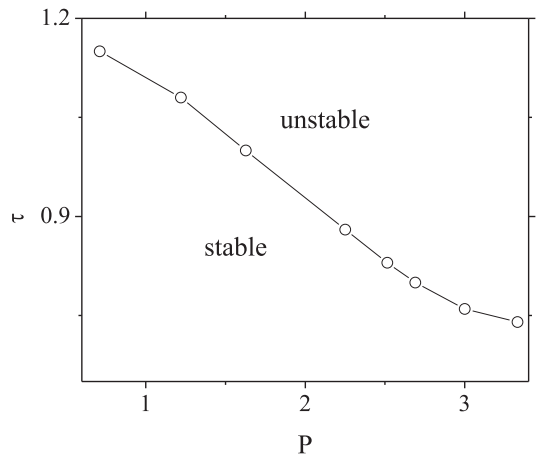} \caption{The
border between stable and unstable on-site stationary solitons in
the parameter plan of $\left( \protect\tau ,P\right) $, for
coupling parameter $C=0.8$. The instability criterion is
$\mathrm{Re(EV)}>10^{-3}$.} \label{fig7}
\end{figure}

The critical value of $\tau $ at which the instability of the on-site
solitons sets in varies as a function of the soliton's norm, as shown in
Fig. \ref{fig7}. The delay effects, including the corresponding instability,
stronger affect tightly localized modes with a large norm, whose behavior is
dominated by the nonlinearity, then their broad small-amplitude
counterparts. However, it is necessary to mention that tightly pinned modes
(essentially, those sitting on a single lattice site) may be actually
unphysical, if the underlying continual model, approximated by the discrete
one, admits the collapse, as in the case of the one-dimensional
nonpolynomial Schr\"{o}dinger equation which describes the Bose-Einstein
condensate with the intrinsic attraction loaded into a deep optical-lattice
potential \cite{Luca}.

\begin{figure}[tbp]
\includegraphics [width=5cm]
{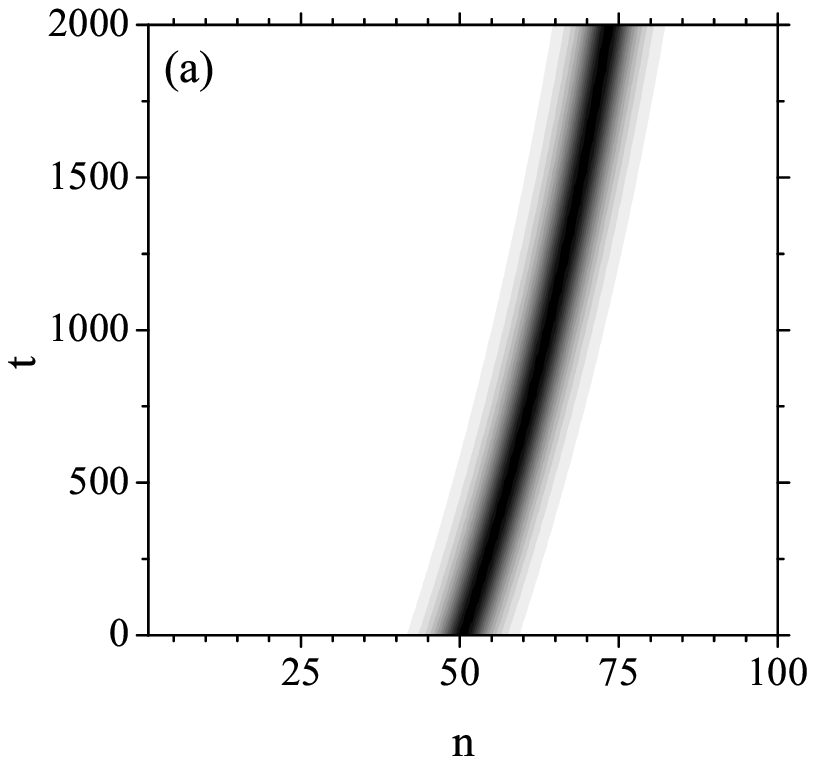}\includegraphics [width=5cm]{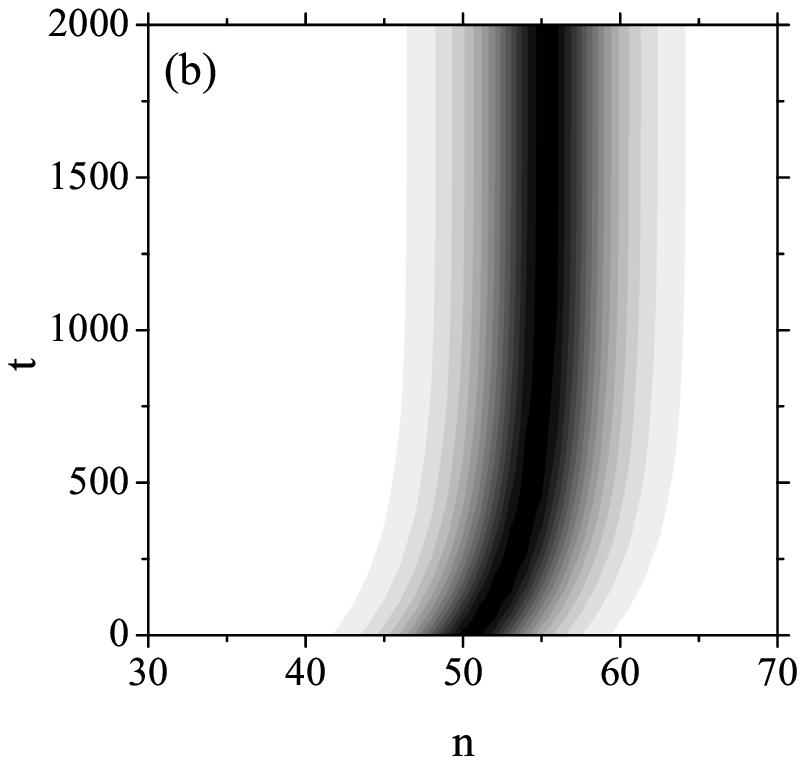}%
\includegraphics
[width=5cm]{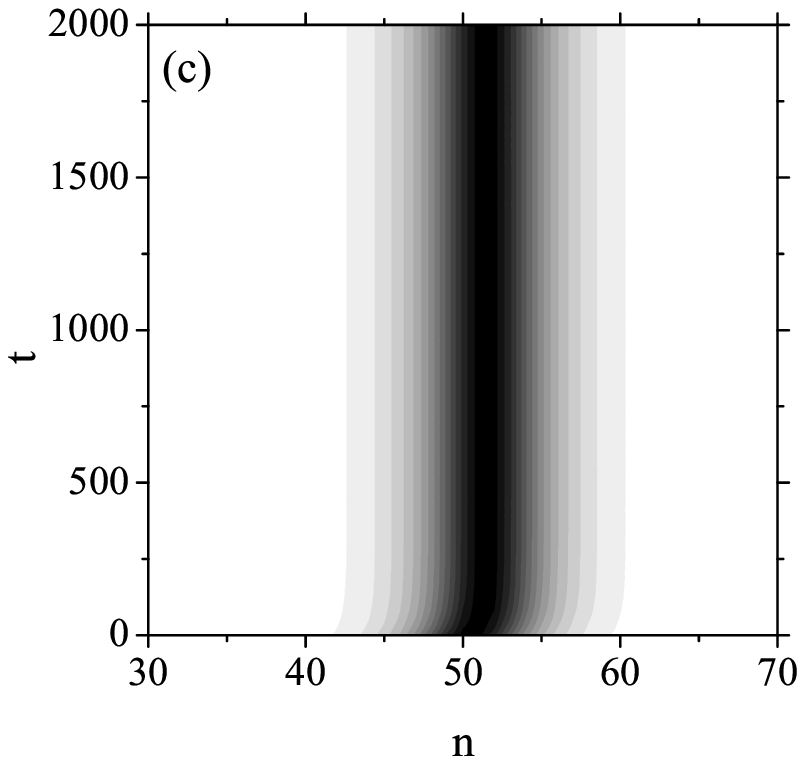}
\caption{The evolution of a kicked inter-site soliton with $\protect\mu %
=-0.12$ and $P=1.22$ in the lattice with time-delayed
nonlinearity. The kick's strength is $\protect\alpha =0.01$, and
the lattice coupling constant is $C=0.8$. The analysis has shown
that this soliton is marginally stable, in its static form, for
$\protect\tau =0.01$ and $\protect\tau =0.1$, and unstable for
$\protect\tau =0.5$. These values of $\protect\tau $ correspond,
respectively, to panels (a), (b), and (c). In the first case (very
small $\protect\tau $), the kicked soliton keeps moving freely. In
the second case (moderately small $\protect\tau $), a moving
breather is formed, which gradually comes to a halt (b). Large
$\protect\tau $ makes the soliton completely immobile (c).}
\label{fig8}
\end{figure}

Local modes which are members of marginally stable inter-site and on-site
soliton pairs with equal norms, which exist in the case of small $\tau $,
can be set in motion across the lattice by the application of the kick,
i.e., taking $\psi _{n}(t=0)=\phi _{n}\exp {(i\alpha n)}$, similar to the
behavior of the marginally stable solitons in the lattice with the
instantaneous nonlinearity. Therefore, the concept of the effective
Peierls-Nabarro barrier can be applied to the lattices with the time-delayed
nonlinearity. However, larger values of $\tau $ enhance the temporal
correlation between time-distant events, favoring bringing moving solitons
to a halt, as shown in Fig. \ref{fig8}.

\section{Conclusion}

In this work, we aimed to study the stability and dynamics of unstaggered
localized modes in nonlinear lattices with the delayed nonlinear response.
The discrete solitons are destroyed when the delay time increases above a
certain critical value, in agreement with the observation reported in Ref.
\cite{moura} that the wave-packet delocalization, implying a breakdown of
the self-trapping, is a delay-induced transition.

In the case of the fast nonlinear response (short temporal delay), the
exponential-instability growth rate of inter-site solitons decreases, in
comparison with the instantaneous model. However, full stabilization of the
inter-site solitons does not occur. Marginally stable inter-site solitons,
which feature a virtually vanishing real instability EV\ (eigenvalue), and
find their on-site counterparts with equal norms, have been found in the
case of a small delay time ($\tau $), as well as for larger $\tau $ near $%
\mu =0$ (i.e., for broad solitons). Under the action of perturbations, these
solitons evolve into breathing localized modes. The marginally stable
solitons can be transformed into moving localized modes by the application
of a kick.

At large values of $\tau $, all solitons, of both the on-site and inter-site
types, are unstable. In direct simulations, they decay into diffractive
patterns. In fact, this means that the localized mode spreads out before the
nonlinearity commences to act in the slowly responding media. The transition
to this instability is indicated by the expansion of the EV spectrum for
small perturbations around the discrete soliton, with the emergence of a
growing number of complex EVs with significant positive real parts.

The results reported in this work clearly demonstrate that the dynamical
properties of localized modes in self-focusing media are essentially
affected by the temporal delay of the nonlinearity. These findings
complement the recently published study of effects of the time-delayed
nonlinear response on the transition to delocalization of wave packets \cite%
{moura}. Altogether, these results provide for an insight into the dynamics
of localized patterns in slowly responding media.

A.M. and Lj.H. acknowledge support from the Ministry of Science, Serbia
(project 141034). The work of B.A.M. was supported, in a part, by grant No.
149/2006 from the German-Israel Foundation.

\end{document}